\documentclass[12pt]{article}

\def\Tr{{\rm Tr}}

\def\I{{\cal I}}

\def\S{{\cal S}}

\def\s{{\bf s}}

\def\half{\frac{1}{2}}
\def\RR{R-R }
\def\NSNS{NS-NS }
\def\NSNST{NS-NS,T}
\def\RRT{R-R,T}

\def\spc{\;\;\;\;}

\def\dh{$\hat{\mbox{D}}$}

\def\Zop{\bbbz} 
\def\Rop{\bbbr}   

\def\pmb#1{\setbox0=\hbox{#1}
 \kern-.025em\copy0\kern-\wd0
 \kern.05em\copy0\kern-\wd0
 \kern-.025em\raise.0433em\box0 }

\def\dh{$\hat{\mbox{D}}$}

\def\3{\ss}
\def\sq{\hbox{\rlap{$\sqcap$}$\sqcup$}}
\def\qed{\ifmmode\sq\else{\unskip\nobreak\hfil
\penalty50\hskip1em\null\nobreak\hfil\sq
\parfillskip=0pt\finalhyphendemerits=0\endgraf}\fi}

\def\bbbz {{\sf Z\!\!Z}}

\def\bbbr {{\rm I\!R}}

\newcommand{\ket}[1]{|#1\rangle}

\def\b,#1,#2(#3){\left|B#1,#2\right>_{#3}}

\def\gso#1,#2{\frac{1}{4}(1#1(-1)^F)(1#2(-1)^{{\tilde F}})}

\def\xxx#1           {{\sf hep-th/#1} }

\def\npb#1(#2)#3     {Nucl. Phys. {\bf B#1} (#2) #3 }
\def\rep#1(#2)#3     {Phys. Rept.{\bf #1} (#2) #3 }
\def\pla#1(#2)#3     {Phys. Lett. {\bf #1A} (#2) #3 }   
\def\plb#1(#2)#3     {Phys. Lett. {\bf #1B} (#2) #3 }
\def\prl#1(#2)#3     {Phys. Rev. Lett.{\bf #1} (#2) #3 }
\def\prd#1(#2)#3     {Phys. Rev. {\bf D#1} (#2) #3 }
\def\ap#1(#2)#3      {Ann. Phys. {\bf #1} (#2) #3 }
\def\rmp#1(#2)#3     {Rev. Mod. Phys. {\bf #1} (#2) #3 }
\def\cmp#1(#2)#3     {Comm. Math. Phys. {\bf #1} (#2) #3 }
\def\mpla#1(#2)#3    {Mod. Phys. Lett. {\bf A#1} (#2) #3 }
\def\ijmp#1(#2)#3    {Int. J. Mod. Phys. {\bf A#1} (#2) #3 }
\def\cqg#1(#2)#3     {Class. Quant. Grav. {\bf #1} (#2) #3 }
\def\am#1(#2)#3      {Adv. Math. {\bf #1} (#2) #3 }
\def\im#1(#2)#3      {Invent. Math. {\bf #1} (#2) #3 }
\def\jhep#1(#2)#3    {JHEP {\bf #1}(#2) #3 }
\def\npps#1(#2)#3    {Nucl.Phys.Proc.Suppl. {\bf #1}(#2) #3 }
\def\jgp#1(#2)#3     {J. Geom. Phys. {\bf #1}(#2) #3 }

\def\dstyler#1       {\documentstyle{report}[#1]}
\def\dstylea#1       {\documentstyle{article}[#1]}
\def\bd              {\begin{document}}
\def\ed              {\end{document}}
\def\be	             {\begin{equation}}
\def\ee              {\end{equation}}
\def\ba              {\begin{eqnarray}}
\def\ea              {\end{eqnarray}}
\def\ni              {\noindent}
\def\bb#1            {\begin{thebibliography}{#1}}
\def\eb              {\end{thebibliography}}

\def\etal {{\em et al.} }
\def\w    {\;_\wedge}
\def\ie   {{\em i.e.}}
\def\ibid {{\em ibid.}}

\advance\voffset by -1.5cm
\advance\hoffset by -2.1cm
\begin{document}

\thispagestyle{empty}
\def\thefootnote{\fnsymbol{footnote}}
\begin{flushright}
  hep-th/0005153\\
  DAMTP-2000-37
 \end{flushright}
\vskip 0.5cm

\begin{center}\LARGE
{\bf Dirichlet Branes on a Calabi-Yau Three-fold Orbifold}
\end{center}
\vskip 1.0cm
\begin{center}
{\large  Bogdan Stefa\'nski, jr.\footnote{E-mail  address: 
{\tt B.Stefanski@damtp.cam.ac.uk}}} 

\vskip 0.5 cm
{\it Department of Applied Mathematics and Theoretical Physics \\
Centre for Mathematical Sciences \\
Wilberforce Road \\
Cambridge CB3 0WA, U.K.}
\end{center}

\vskip 1.0cm

\begin{center}
March 2000
\end{center}

\vskip 1.0cm

\begin{abstract}
The D-brane spectrum of a $\Zop_2\times\Zop_2$ Calabi-Yau three-fold
orbifold of toroidally
compactified Type IIA and Type IIB string theory is analysed
systematically. The corresponding K-theory groups are determined 
and complete agreement is found. New kinds of stable non-BPS D-branes are
found, whose stability regions are far more complicated than the
previously discussed non-BPS D-branes. The decay channels of 
non-BPS D-branes beyond their stability regions are
identified. Finally the T-dual orbifold is
analysed and a suitable K-theory is found. 
\end{abstract}

\vskip 1.0cm
\begin{center}
 PACS 11.25.-w, 11.25.Sq
\end{center}

\vfill
\setcounter{footnote}{0}
\def\thefootnote{\arabic{footnote}}
\newpage

\renewcommand{\theequation}{\thesection.\arabic
{equation}}

\section{Introduction}
\setcounter{equation}{0}

\indent The study of non-BPS D-branes\footnote{For some recent reviews
see~\cite{SenRev,lr,sch}.} has provided a better understanding of
several aspects of string theory. The classification of D-branes in
terms of K-theory~\cite{MM,WittenK,Horava} can be physically
understood by thinking of D-branes as solitonic solutions of the
tachyon field of higher dimensional brane-anti-brane 
systems~\cite{sen2,Sen3,WittenK,Sen10}. The study of the tachyon
potential using string field theory~\cite{SenZw2, berk,BSZ,mt}
has given strong evidence to the
conjecture that a brane-anti-brane system annihilates into the
vacuum. Due to the presence of a single decay mode unstable non-BPS
D-branes have been interpreted as sphaleron solutions of string
theory~\cite{hhk} and some understanding of bosonic D-branes~\cite{hk}
has now also been developed. The confinement process of the unbroken
gauge group boson on the world-volume of annihilating brane-anti-brane
systems~\cite{srednicki,WittenK} has been better
understood~\cite{Yi,Sen9,bhy}. 

\indent Another aspect of recent developments has involved the study of stable
non-BPS D-branes through the boundary state formalism~\cite{CLNY,PolCai}.  
This was first considered in a specific context in~\cite{sen2,BG2}. Over 
the past year the boundary state formalism has been used to 
construct many more examples of non-BPS 
D-branes~\cite{Sen6,BG3,Gukov,Frau2,DG,GabSen,S3,mot,rs,mst,ep}.
These constructions have tested the connection between D-branes and
K-theory~\cite{BGH,Gukov,S3} as well as testing S-dualities beyond the
BPS constraints. In particular the dualities between
heterotic and Type II theories~\cite{BG3} and heterotic and Type I
theories~\cite{Sen1,S4,gls} have been investigated. Due to the fact
that certain non-BPS D-branes are the lightest states carrying a
particular charge it has been possible to identify their duals and
compare regions of stability~\cite{S4} as well as interaction
properties~\cite{gls} in the dual theories. 

\indent The integrally (rather than torsion) charged non-BPS D-branes have so
far been constructed on $\Zop_2$
orbifolds~\footnote{In~\cite{Sen6,mot} this had been generalised
somewhat to a Calabi-Yau
$\Zop_2\times\Zop_2^\prime$ orbifold, where $\Zop_2^\prime$ is freely acting.}
and have been interpreted in the blow-up~\cite{Sen6} as coming from
certain volume minimising non-supersymmetric cycles of the manifold
which in the orbifold limit shrink to one-cycles. Here we discuss a
$\Zop_2\times\Zop_2$ orbifold of Type IIA and IIB. In 
particular we
study BPS and non-BPS D-branes on such an orbifold.
The orbifold
is a particular limit of a Calabi-Yau three-fold, and as such
preserves N=2, D=4 supersymmetry. The action of the two generators
$g_1$ and $g_2$ will be taken as
\ba
g_1(x^0,\dots,x^9)&=&(x^0,x^1,x^2,x^3,x^4,-x^5,-x^6,-x^7,-x^8,x^9)\,\\
g_2(x^0,\dots,x^9)&=&(x^0,x^1,x^2,-x^3,-x^4,x^5,x^6,-x^7,-x^8,x^9)\,
\ea
and we label $g_3=g_1g_2$. Thus both $\Zop_2$'s have fixed
points. This orbifold has been studied
previously~\cite{vw94,MD,BGe,MR,DF}. In particular in~\cite{MD}
fractional D-branes on it were discussed and it was noted that there
are two kinds of fractional D-branes, those that live at the fixed
hyperplanes of one of the $g_i$ (we shall refer to these as {\em singly}
fractional D-branes) and those that live at the fixed hyperplanes of
all $g_i$ (we shall refer to these as {\em totally} fractional D-branes). 

\indent In this paper we use the boundary
state formalism to investigate the full D-brane spectrum, BPS and
non-BPS on this orbifold and study the stability regions of non-BPS
\dh-branes on the orbifold. We give a boundary state description of both
kinds of fractional D-branes as well as bulk D-branes wrapping the
special Lagrangian three-cycle of the Calabi-Yau orbifold. 
We find non-BPS \dh-branes similar to
those of~\cite{S3}. We also find that there are new kinds of non-BPS
\dh-branes on this orbifold, and we refer to these too as {\em truncated}
\dh-branes~\cite{S3}. The truncated \dh-branes break all
supersymmetry and are charged under twisted \RR fields. 
The relevant K-theory groups are evaluated and 
complete agreement between the D-branes we have constructed and
K-theory is found.

\indent Some of the new \dh-branes have very unusual
domains of stability and besides coupling to twisted \RR
sectors they couple to certain twisted \NSNS sectors as well. We
investigate the decay channels of the various \dh-branes. Besides the
usual decays into brane-anti-brane pairs of BPS fractional D-branes
there are other decay channels in which non-BPS \dh-branes decay into
one another. There are also certain decay channels whose decay
products are unknown. When the decay products are known the masses and
charges of the relevant objects are the same at critical radii,
indicating that the process is a marginal deformation in the conformal
field theory~\cite{Sen3,Sen4,MajSen2}.

\indent The construction of boundary states describing the
D-branes is given in section 2, as well as the appendices.
In section 3 we compute the relevant equivariant K-theory groups and
find complete agreement with section 2. In section 4 we consider the
compact orbifold and we discuss in 
detail the minimal charge configurations of the D-branes.
The stability regions of the truncated non-BPS \dh-branes are analysed in 
section 5. The new kind of truncated \dh-brane
exhibits some remarkable stability properties as given in
equation~(\ref{funnybound}). In section 6 we identify most of the
decay channels for the various \dh-branes. Finally section 7 discusses
D-branes on the T-dual orbifold. A new K-theory group suitable for
such an orbifold is defined.

\section{Non-compact orbifold}\label{sec2}
\setcounter{equation}{0}

In this section we discuss the non-compact orbifold, while later on
the directions $x^3,\dots,x^8$ will be compactified on circles. We 
first briefly discuss the massless spectrum, and then construct 
relevant D-brane boundary states.

The $\Zop_2\times\Zop_2$ orbifold discussed here has been studied in
the past~\cite{vw94,MD}. In particular its closed string spectrum has
been worked out in some detail~\cite{vw94}. Here we point out some of
the points of present importance. Each $g_i$,
$i=1,2,3$ gives rise to a twisted sector. Both the
twisted NS and R sectors are massless and have zero modes, those of the R
sector are in the directions unaffected by $g_i$, while the zero modes
in the NS sector are in the directions inverted by $g_i$. The lowest
lying states in the twisted \RR sector transform as a tensor product
of $SO(4)$ (which contains the spacetime SO(2)), and are further
required to be invariant under the remaining orbifold projections. 
The twisted \NSNS sector ground states transform as bispinors of an
internal $SO(4)$, and have to be invariant under the other elements of
the orbifold group. They give rise to four dimensional scalars. 
We denote the $g_i$-twisted sectors by  \NSNST$g_i$ and \RRT$g_i$.

Boundary states corresponding to physical
D-branes consist of linear combinations of boundary states
from the various closed
string sectors which are GSO and orbifold invariant.
We refer to a D-brane as of type $(r;{\bf s})=(r;s_1,s_2,s_3)$, if it is a
$r+s_1+s_2+s_3$-brane, and extends along $r+1$, $s_1$, $s_2$, $s_3$ of the
directions $(x^0,\dots,x^2,x^9)$, $(x^3,x^4$), $(x^5,x^6)$, $(x^7,x^8)$,
respectively.\footnote{With this convention $g_i$ acts trivially on
the $s_i$ directions.} Due to the
presence of fermionic zero modes in the various sectors
requiring GSO and orbifold invariance places restrictions
on $r$ and the $s_i$. These conditions are analysed in Appendix~\ref{appb}.

A boundary state which corresponds to a physical D-brane has to satisfy
certain consistency criteria. In particular the string with endpoints on
the D-brane has to be a suitably projected open string. It turns out that
in the $\Zop_2\times\Zop_2$ orbifold there are four kinds of combinations
of boundary states from the various closed string sectors which correspond
to physical D-branes. These include a bulk D-brane
\be
\ket{B(r;{\bf s})}_{\mbox{\scriptsize\NSNS}}+
\varepsilon\ket{B(r;{\bf s})}_{\mbox{\scriptsize\RR}}\,,
\ee
where $\varepsilon=\pm 1$, and a fractional D-brane
\be
\ket{B(r{\bf s};)}_{\mbox{\scriptsize\NSNS}}+\varepsilon
\ket{B(r;{\bf s})}_{\mbox{\scriptsize\RR}} 
+\sum_{i=1}^3\varepsilon_i\left(
\ket{B(r;{\bf s})}_{\mbox{\scriptsize\NSNST$g_i$}}+
\varepsilon\ket{B(r;{\bf s})}_{\mbox{\scriptsize\RRT$g_i$}}\right)\,,
\label{fract}
\ee
for which $\varepsilon_i\varepsilon_j=\varepsilon_k$ for $i\neq j\neq k\neq i$;
both these objects are BPS as can be seen for example from the vanishing
of the cylinder amplitudes computed in appendix~\ref{appa}. In the above
$\varepsilon,\varepsilon_i$ indicate the sign of the various \RR
charges. Further there are
truncated non-BPS \dh-branes charged under one type of twisted \RR field
\be
\ket{B(r;{\bf s})}_{\mbox{\scriptsize\NSNS}}+\varepsilon
\ket{B(r;\s)}_{\mbox{\scriptsize\RRT$g_i$}} ,\spc i=1,2,3\,. 
\label{sgtr}
\ee
Their boundary state is very similar to that of the \dh-branes
in~\cite{S3}. Finally there is a second type of truncated non-BPS
\dh-brane whose boundary state is
\ba
\ket{B(r;\s)}_{\mbox{\scriptsize\NSNS}}+\varepsilon_i
\ket{B(r;\s)}_{\mbox{\scriptsize\RRT$g_i$}} 
+\varepsilon_i\varepsilon_j
\ket{B(r;\s)}_{\mbox{\scriptsize\NSNST$g_k$}}+\varepsilon_j
\ket{B(r;\s)}_{\mbox{\scriptsize\RRT$g_j$}}\,,
\nonumber \\
\spc i,j,k=1,2,3,\spc
\mbox{s.t. }\epsilon_{ijk}\neq 0 \,.
\label{dbtr}
\ea 
This is a new kind of \dh-brane which couples to a twisted \NSNS
sector as well as to twisted \RR sectors.

A second consistency condition is that a string beginning on any one of the
above branes and ending on a different one must describe an open string.
In~\cite{S3} this condition ensured that for any given $(r,s)$ if a
fractional and truncated $(r,s)$-branes existed then the fractional brane
would be of smaller mass and charge rendering the truncated brane
unstable. Applying the condition to the branes above we again conclude
that if a fractional and truncated brane exist for some $(r;\s)$
then it is the fractional object that is minimal and stable. Similarly
one shows that if for some $(r;\s)$ both truncated objects exist
it is the one in equation~(\ref{dbtr}) that is minimally charged and so is
fundamental. 

Given these consistency conditions the D-brane spectrum can then be
determined. The detailed analysis can be found in appendix~\ref{appb}. Bulk and
fractional D-branes exist for $(r;\s)$ of the form
\be
(r;0,0,0)\,,(r;0,0,2)\,,(r;0,2,0)\,,(r;2,0,0)\,,
(r;0,2,2)\,,(r;2,0,2)\,,(r;2,2,0)\,,(r;2,2,2)\,,
\label{frbr}
\ee
where $r$ is even/odd for Type IIA/IIB.  The elementary objects above
are {\em totally} fractional
branes which live on the fixed point of the whole orbifold group
($x^3=\dots=x^8=0$) and are charged under all three twisted \RR
sectors. The corresponding K-theory group should be
$\Zop^{\oplus 4}$. The totally fractional branes correspond to a
single brane in the covering space. Two
such D-branes with the same bulk and $g_i$-twisted \RR charges but
opposite remaining charges can come together to form a {\em singly}
fractional brane, which is stuck to the fixed plane of
$g_i$. Two singly fractional branes with opposite twisted \RR
charge can come together and form a bulk brane. Further bulk D-branes exist
for $(r;1,1,1)$. Since these are charged under the untwisted \RR field,
the corresponding K-group should be $\Zop$.

Truncated \dh-branes with a boundary state given by equation~(\ref{sgtr})  
exist for $(r;\s)$ of the form\footnote{We disregard here the
tachyon that arises when considering decompactified $s_i>0$ truncated
branes~\cite{S3}, as on compactification the D-brane will stabilise for certain
values of the compactification radii.}
\be
(r;0,1,1),\;(r;1,0,1),\;(r;1,1,0),\;
(r;2,1,1),\;(r;1,2,1),\;(r;1,1,2)\,,\label{scharge}
\ee
with $r$ even/odd for Type IIA/IIB.
These \dh-branes are stuck at the fixed points of the $g_i$ under
which they are charged, and one expects the corresponding K-group to
be $\Zop$. Truncated \dh-branes with a boundary state
given by equation~(\ref{dbtr}) exist for $(r;\s)$ of the form
\ba
(r;0,0,1),\;(r;1,0,0),\;(r;0,1,0),\;
(r;2,0,1),\;(r;1,2,0),\;(r;2,1,0),\nonumber \\
(r;0,2,1),\;(r;1,0,2),\;(r;0,1,2),\;
(r;2,2,1),\;(r;1,2,2),\;(r;2,1,2)\,.\label{dcharge}
\ea
Here $r$ is even/odd for Type IIA/IIB. 
The basic such branes are stuck at the fixed points of the whole
orbifold group and are charged under two twisted \RR fields suggesting
that the K-group should be $\Zop\oplus\Zop$. However, as with the
fractional branes there are also
\dh-branes with the above $(r;\s)$ which are charged under
only $g_i$-twisted \RR fields and as such are only stuck to the fixed
points of $g_i$.

\section{K-theory analysis in uncompactified theory}\label{sec3}
\setcounter{equation}{0}

The K-groups relevant to this orbifold are the
$\Zop_2\times\Zop_2$-equivariant K-groups~\cite{Segal} with compact
support
\be
K^*_{\Zop_2\times\Zop_2}(\Rop^{a;b,c,d})\,,
\ee
where $a=0,\dots,4$ and $b,c,d=0,1,2$. In the above the directions
$a$ are left invariant by $G=\Zop_2\times\Zop_2$, $c$ and $d$ are
inverted by the first $\Zop_2$ and $b,c$ are inverted by the second
$\Zop_2$. These groups exhibit
complex Bott periodicity in $a,b,c$ and $d$  thus the answer depends
only on the parity of $a,b,c$ and $d$. Further there is a symmetry
between $b,c,d$. For example
$K^*_G(\Rop^{a;b,c,d})=K^*_G(\Rop^{a;c,b,d})$. As a result we need only
compute very few terms. Since the representation ring of
$\Zop_2\times\Zop_2$ is $\Zop^{\oplus 4}$ we have
\be
K^*_G(\Rop^{a;b,c,d})=\left\{\begin{array}{cc}\Zop^{\oplus
4} & a+* \mbox{ even} \\0& a+* \mbox{ odd}\end{array}\right. 
\ee
for $b,c,d$ even.
Fractional branes in the decompactified theory can be
charged under three twisted \RR fields as well as the untwisted \RR
field, the above K-groups confirm the presence of all fractional branes. 
Since $g_1$ acts trivially on $\Rop^{0;1,0,0}$ we have
\be
K^*_G(\Rop^{0;1,0,0})=K^*_{\Zop_2}(\Rop^{0,1})\otimes
R[\Zop_2]\,,
\ee
where on the right hand side $\Zop_2$ inverts the line
$\Rop^{0,1}$. The right-hand side groups have been obtained
in~\cite{Gukov,S3}. Thus we have for $c,d$ even
\be
K^*_G(\Rop^{a;1,c,d})=\left\{\begin{array}{cc}\Zop^{\oplus
2} & a+* \mbox{ even} \\0& a+* \mbox{ odd}\end{array}\right. 
\ee
Similar results hold for permutations of $1,c,d$. This is in agreement
with the presence of truncated \dh-branes 
in~(\ref{dcharge}). In order to compute 
$K^*_G(\Rop^{0;1,1,2k})$ we re-write it as
\be
K^*_G(\Rop^{0;1,1,0})=K^*_G(\Rop^{0;1,0,0}\times 
D^1,\Rop^{a;1,0,0}\times S^0)\,,
\ee
where $S^0\subset D^1\subset\Rop^{0;0,1,0}$ are the zero-sphere (two
points) and the one-disc (the interval). By homotopy equivalence
we have
\be
K^*_G(X\times D^1)=K^*_G(X)\,,
\label{hom}
\ee
for any $X$ and
\be
K^*_G(\Rop^{0;1,0,0}\times S^0)=K^*_{\Zop_2}(\Rop^{0,1})\,,
\ee
where on the right hand side $\Zop_2$ inverts the real line
$\Rop^{0,1}$. We may now use the long exact sequence
\ba
\cdots&\rightarrow&\!\! K^{-1}_{G}(\Rop^{0;1,0,0}\times S^0) 
\rightarrow  K_{G}(\Rop^{0;1,0,0}\times \Rop^{0;0,1,0}) \;
\rightarrow \; K_{G}(\Rop^{0;1,0,0}\times D^1)\nonumber \\
& \rightarrow &\!\! K_{G}(\Rop^{0;1,0,0}\times S^0) \;
\rightarrow \; K^1_{G}(\Rop^{0;1,0,0}\times \Rop^{0;0,1,0}) 
\rightarrow \; K^1_G(\Rop^{0;1,0,0}\times D^1) \;
\rightarrow\cdots
\ea
\ni which in turn becomes
\be
0\rightarrow K_{G}(\Rop^{0;1,1,0})
 \rightarrow \Zop\oplus\Zop\rightarrow \Zop
 \rightarrow K^1
_{G}(\Rop^{0;1,1,0})\rightarrow 0\,.
\ee
It is not difficult to see that the map
$\Zop\oplus\Zop\rightarrow\Zop$ is surjective. This gives for $d$ even
\be
K^*_G(\Rop^{a;1,1,d})=\left\{\begin{array}{cc}\Zop
 & a+* \mbox{ even} \\0& a+* \mbox{ odd}\end{array}\right. 
\ee
Together with the permutations of $1,1,d$ this confirms the presence
of truncated \dh-branes in equation~(\ref{scharge}).
Finally we have
\be
K^*_{G}(\Rop^{0;1,1,1})=K^*_{G}(\Rop^{0;1,1,0}\times
D^1,\Rop^{a;1,1,0}\times S^0)\,,
\ee
\ni where $S^0\subset D^1\subset\Rop^{0;0,0,1}$. Using
equation~(\ref{hom}) the results above and
\be
K_G^*(\Rop^{0;1,1,0}\times\S^0)=K_{\Zop_2}^*(\Rop^{0,2})\,,
\ee
the following sequence is exact
\be
0\rightarrow K_{G}(\Rop^{0;1,1,1})
 \rightarrow \Zop\rightarrow \Zop\oplus\Zop
 \rightarrow K^1_{G}(\Rop^{0;1,1,1})\rightarrow 0\,.
\ee
One can show using this exact sequence that~\cite{Segal1}
\be
K^*_G(\Rop^{a;1,1,1})=\left\{\begin{array}{cc}\Zop
 & a+* \mbox{ odd} \\0& a+* \mbox{ even}\end{array}\right. \\ 
\ee
This confirms the presence of bulk BPS branes with $s_1=s_2=s_3=1$.
We have demonstrated complete agreement between K-theory and
D-branes on this orbifold.

\section{The compactified orbifold}
\setcounter{equation}{0}

From now on we take the directions $x^3,\dots,x^8$ to be
compactified on circles of radii $R_3,\dots,R_8$. This introduces new
fixed points at $x^i=\pi R_i$ and gives rise to new twisted sectors - in
total 48 twisted \RR sectors and 48 twisted \NSNS sectors. D-branes with
$s_i>0$ will be charged under the twisted sectors over which they
stretch. The structure of the boundary states and in particular the
normalisation is described in appendix~\ref{appa}.  The allowed
charges for various
D-branes are restricted by factorisation. In particular a brane
charged under many \RR charges cannot have an arbitrary choice of
minimal positive and negative charges. This was already encountered
in~\cite{S3} where for example a \dh$(r,2)$-brane in the $\I_n(-1)^{F_l}$
orbifolds
was charged under four twisted \RR fields but the minimally charged
branes had to have an even number of negative charges. This behaviour
is typical and we encounter it here as well. The restriction arises
as a result of cylinder-annulus consistency. The process of checking
what sign freedom one has in the closed string sector in order to
factorise on a consistent open string amplitude is laborious, here we
summarise the results. 

In the following subsections we shall discuss the D-brane spectrum of
the compactified orbifold, paying attention to the allowed charges,
as well as stability regions and decay products of the non-BPS branes.
For simplicity we concentrate on the fractional  D$(r;0,0,0)$ and
D$(r;0,0,2)$-branes and truncated \dh$(r;0,0,1)$ and
\dh$(r;0,1,1)$-branes. The extension of these results to the other
branes is obvious.

\subsection{The D$(r;0,0,0)$-brane}

The fully fractional D$(r;0,0,0)$-brane's consistent boundary state is given by
\ba
\ket{D(r;0,0,0),\varepsilon,\varepsilon_i}&=&
\ket{D(r;0,0,0)}_{\mbox{\scriptsize  
\NSNS}}+\varepsilon\ket{D(r;0,0,0)}_{\mbox{\scriptsize \RR}}\nonumber \\
& &\spc+\;
\sum_{i=1}^3 
\varepsilon_{i}\left(\ket{D(r;0,0,0)}_{\mbox{\scriptsize
\NSNST$g_i$}}+\varepsilon\ket{D(r;0,0,0)}_{\mbox{\scriptsize
\RRT$g_i$}}\right)\,,\label{eq41}
\ea
with $\varepsilon_3=\varepsilon_1\varepsilon_2=\pm 1$. If we denote by
$[a;b,c,d]$ the four charges under which the D-brane is charged (with
$a$ corresponding to the bulk charge, and $b,c,d$ to the twisted charges)
the allowed configurations of minimal charge are
\ba
[1;1,1,1],\;[1;1,-1,-1],\;[1;-1,1,-1],\;[1;-1,-1,1],\nonumber \\
\,[-1;-1,-1,-1],\;[-1;-1,1,1],\;[-1;1,-1,1],\;[-1;1,1,-1]\,.
\ea
From this it is easy to see that a singly fractional brane is a sum of two
totally fractional branes. For example
$[2;0,2,0]=[1;1,1,1]+[1;-1,1,-1]$ or $[-2;0,2,0]=[-1;-1,1,1]+[-1;1,1,-1]$.

\subsection{The D$(r;0,0,2)$-brane}

We take the fully fractional D$(r;0,0,2)$-brane to
extend along $x^7$ and $x^8$ and to be fixed at
$x^3=\dots=x^6=0$. Its' boundary state is
\ba
\ket{D(r;0,0,2),\varepsilon,\varepsilon_i,\theta_j}\!\!&=&\!\!\!\sum_{w_7,w_8}
e^{i(\theta_7 w_7+\theta_8 w_8)}(\ket{D(r;0,0,2),w_7,w_8}_{\mbox{\scriptsize
\NSNS}}+\varepsilon\ket{D(r;0,0,2),w_7,w_8}_{\mbox{\scriptsize \RR}})
\nonumber \\ & &
+\sum_{i\in\{1,2\}}
\left(\varepsilon_i(\ket{D(r;0,0,2)}_{\mbox{\scriptsize
\NSNST$g_{i,1}$}}+\varepsilon\ket{D(r;0,0,2)}_{\mbox{\scriptsize
\RRT$g_{i,1}$}})\right.
\nonumber \\ & &\spc
+\varepsilon_ie^{i\theta_7}(\ket{D(r;0,0,2)}_{\mbox{\scriptsize
\NSNST$g_{i,2}$}}+\varepsilon\ket{D(r;0,0,2)}_{\mbox{\scriptsize
\RRT$g_{i,2}$}})
\nonumber \\ & &\spc
+\varepsilon_ie^{i\theta_8}(\ket{D(r;0,0,2)}_{\mbox{\scriptsize
\NSNST$g_{i,3}$}}+\varepsilon\ket{D(r;0,0,2)}_{\mbox{\scriptsize
\RRT$g_{i,3}$}})
\nonumber \\ & &\left.\spc
+\varepsilon_ie^{i(\theta_7+\theta_8)}(\ket{D(r;0,0,2)}_{\mbox{\scriptsize
\NSNST$g_{i,4}$}}+\varepsilon\ket{D(r;0,0,2)}_{\mbox{\scriptsize
\RRT$g_{i,4}$}})\right)
\nonumber \\ & &\spc
+\varepsilon_1\varepsilon_2
\sum_{w_7,w_8}e^{i(\theta_7 w_7+\theta_8 w_8)}
(\ket{D(r;0,0,2),w_7,w_8}_{\mbox{\scriptsize
\NSNST$g_3$}}
\nonumber \\ & &\spc\spc\spc\spc\spc
+\varepsilon\ket{D(r;0,0,2),w_7,w_8}_{\mbox{\scriptsize
\RRT$g_3$}})\label{eq43}
\ea
From this it is apparent that D$(r;0,0,2)$-branes can only have
an even number of negative \RRT$g_1$ and \RRT$g_2$ charges, and
further that
the sign of  the \RRT$g_3$ charge is fixed by the other twisted
charges. A D$(r;0,0,2)$-brane with all positive \RRT$g_1$ and bulk
\RR charges, but negative remaining charges (take
$\varepsilon=\varepsilon_1=+1$, $\theta_7=\theta_8=0$, $\varepsilon_3=-1$) can
be added to a D$(r;0,0,2)$-brane with all positive charges, to give a
brane which is only charged under bulk \RR and \RRT$g_1$ fields, and
so is a singly fractional brane (it can move off the $x^3=x^4=0$ fixed
plane).

\subsection{The \dh$(r;0,0,1)$-brane}

We consider the \dh$(r;0,0,1)$-brane to stretch along the $x^8$ direction
and to be fixed at $x^3=\dots=x^7=0$.
In order to be consistent with the open string partition function the
\dh$(r;0,0,1)$-brane boundary state must be written in the form
\ba
\ket{{\hat D}(r;0,0,1),\varepsilon_i,\theta}&=&\sum_w
e^{i\theta w}\ket{{\hat D}(r;0,0,1),w}_{\mbox{\scriptsize
\NSNS}}+\varepsilon_1\ket{{\hat D}(r;0,0,1)}_{\mbox{\scriptsize
\RRT$g_{1,1}$}}\nonumber \\ & &\spc
+\varepsilon_1e^{i\theta}\ket{{\hat D}(r;0,0,1)}_{\mbox{\scriptsize
\RRT$g_{1,2}$}}+\varepsilon_2\ket{{\hat D}(r;0,0,1)}_{\mbox{\scriptsize
\RRT$g_{2,1}$}}\nonumber \\ & &\spc
+\varepsilon_2e^{i\theta}\ket{{\hat D}(r;0,0,1)}_{\mbox{\scriptsize
\RRT$g_{2,2}$}}
+\varepsilon_1\varepsilon_2\sum_we^{i\theta
w}\ket{{\hat D}(r;0,0,1),w}_{\mbox{\scriptsize
\NSNST$g_3$}}\,.\nonumber \\
\ea
Denoting by $[a,b;c,d]$ the four charges a \dh$(r;0,0,1)$-brane
carries (two from the $g_1$-twisted sector, say $a,b$ and two from the
$g_2$ twisted sector, say $c,d$), the allowed
\dh$(r;0,0,1)$-branes with minimal charge are
\ba
[1,1;1,1],\;[-1,-1;1,1],\;[1,1;-1,-1],\;[1,-1;1,-1],\nonumber \\
\,[-1,1;1,-1],\;[1,-1;-1,1],\;[-1,1;-1,1],\;[-1,-1;-1,-1]\,.
\ea
A singly charged \dh$(r;0,0,1)$-brane with minimal charge comes from 
$[1,1;1,1]+[1,1;-1,-1]=[2,2;0,0]$ or
$[1,-1;1,-1]+[1,-1,-1,1]=[2,-2;0,0]$.

Since the stability region of this
\dh-brane is so different from the non-BPS branes previously
encountered we discuss this case in detail. 
The partition function for a string with both end-points on this \dh-brane is
\ba
\int\frac{dt}{2t}
& &\!\!\!\!\!\!\!\!\!\!\!\!\!
\Tr_{NS-R}\left(\frac{1+(-1)^Fg_1}{2}\frac{1+(-1)^Fg_3}{2}
e^{-2tH_o}\right) \nonumber \\
& = & \!\frac{V_{r+1}}{4 (2\pi)^{r+1}} 
\int\frac{dt}{2t}\! (2t)^{-(r+1)/2}
\left(\frac{f_3^8({\tilde{q}})-f_2^8({\tilde q})}{f_1^8({\tilde q})}\right)
\sum_{n_8\in\Zop} e^{-2t\pi n_8^2 / R_8^2}
\prod_{m=3}^7\sum_{w_m\in\Zop} e^{-2t\pi w_m^2 R_m^2}
\nonumber \\
& & -  \frac{V_{r+1}}{(2\pi)^{r+1}}\int
\frac{dt}{2t}(2t)^{-(r+1)/2}\frac{f_3^{4}({\tilde q})f_4^4({\tilde q})}
{f_1^4({\tilde q})f_2^4({\tilde q})}
\prod_{m=3,4}\sum_{w_m\in\Zop} e^{-2t\pi w_m^2 R_m^2}
\nonumber \\
& & -  \frac{V_{r+1}}{(2\pi)^{r+1}}\int
\frac{dt}{2t}(2t)^{-(r+1)/2}\frac{f_3^{4}({\tilde q})f_4^4({\tilde q})}
{f_1^4({\tilde q})f_2^4({\tilde q})}
\prod_{m=5,6}\sum_{w_m\in\Zop} e^{-2t\pi w_m^2 R_m^2}
\nonumber \\
& & +  \frac{V_{r+1}}{(2\pi)^{r+1}}\int
\frac{dt}{2t}(2t)^{-(r+1)/2}\frac{f_3^{4}({\tilde q})f_4^4({\tilde q})}
{f_1^4({\tilde q})f_2^4({\tilde q})}
\sum_{w_7\in\Zop} e^{-2t\pi w_7^2 R_7^2}
\sum_{n_8\in\Zop} e^{-2t\pi n_8^2 / R_8^2}
\nonumber \\ &\approx&\!
\frac{V_{r+1}}{4(2\pi)^{r+1}}\int\frac{dt}{2t}(2t)^{-(r+1)/2}{\tilde q}^{-1}
\left({\tilde q}^{2(n_8/R_8)^2}+{\tilde q}^{2(w_7R_7)^2}\!+
\sum_{i=3}^7{\tilde q}^{2(w_iR_i)^2}({\tilde q}^{2(n_8/R_8)^2}\!+
{\tilde q}^{2(w_7R_7)^2})\right.
\nonumber \\ & &\left.+
{\tilde q}^{2(w_3R_3)^2+2(w_5R_5)^2}+{\tilde q}^{2(w_3R_3)^2+2(w_6R_6)^2}
+{\tilde q}^{2(w_4R_4)^2+2(w_5R_5)^2}+{\tilde
q}^{2(w_4R_4)^2+2(w_6R_6)^2}\right)\,.
\ea
We have expanded to relevant order in the last line of the
amplitude. It is then clear that the tachyon cancels
provided\footnote{It is straightforward to see how this generalises to other
\dh$(0;2k,2k^\prime,1)$-branes. For example for a \dh$(0;2,0,1)$-brane
stretching along the directions $x^3,x^4,x^8$ the stability region is
\ba
&&R_7\ge\frac{1}{\sqrt{2}}\,,\spc\spc R_8\le\sqrt{2}\,,\nonumber \\ 
&&\frac{1}{R_3^2}+R_5^2\ge\frac{1}{2}\,,\;\; 
\frac{1}{R_4^2}+R_5^2\ge\frac{1}{2}\,, 
\frac{1}{R_3^2}+R_6^2\ge\frac{1}{2}\,,\;\;
\frac{1}{R_4^2}+R_6^2\ge\frac{1}{2}\,.\nonumber
\ea}
\ba
&&R_7\ge\frac{1}{\sqrt{2}}\,,\spc\spc R_8\le\sqrt{2}\,,\\
&&R_3^2+R_5^2\ge\frac{1}{2}\,,\;\; R_4^2+R_5^2\ge\frac{1}{2}\,,\;\;
R_3^2+R_6^2\ge\frac{1}{2}\,,\;\; R_4^2+R_6^2\ge\frac{1}{2}\,.
\label{funnybound}
\ea
Note that by fixing say $R_3,R_4$ to suitable values $R_5,R_6$ can
take on any value and the truncated D-brane will still be stable. 

In order to analyse the decay products of this \dh-brane we take it to
carry a unit of positive twisted \RR charge
in each of the four sectors under which it is charged (two \RRT$g_1$
and two \RRT$g_2$). As can be seen from the above equations there are
three distinct decay channels: $R_8$ can
increase beyond $\sqrt{2}$, $R_7$ can decrease below $1/\sqrt{2}$, and
the bounds for $R_3,R_4,R_5,R_6$ in equation~(\ref{funnybound}) can be
violated. 

For $R_8>\sqrt{2}$ the \dh$(r;0,0,1)$-brane decays, conserving mass
and charge,
into a pair of totally fractional D$(r;0,0,0)$-branes, with opposite 
bulk and \RRT$g_3$ charge. In the notation of equation~(\ref{eq41}) one
of the fractional D-branes has
$\varepsilon=\varepsilon_1=\varepsilon_2=\varepsilon_3=1$, and the
other has $\varepsilon=\varepsilon_1=-\varepsilon_2=\varepsilon_3=-1$. 
The easiest way to compare the mass and charges of the
original truncated brane with those of the decay products is
by perusal of the relevant normalisation constants computed in
appendix~\ref{appa}. The normalisation
constants of the \dh$(r;0,0,1)$-brane are (cf. equations~(\ref{dbtruncnormun})
and~(\ref{dbtruncnormtw}))
\ba
{\cal N}^2_{td,U}&=&\frac{V_{r+1}}{(2\pi)^{r+1}} {1\over 64}
{R_8\over\prod_{m=3}^7 R_m} \,,\\
{\cal N}^2_{td,Tg_1,i}&=&\frac{V_{r+1}}{(2\pi)^{r+1}} {2^3 \over 64}
{1\over R_3 R_4}\,,\spc i=1,2\,, \\
{\cal N}^2_{td,Tg_3,i}&=&\frac{V_{r+1}}{(2\pi)^{r+1}} {2^3 \over 64}
{1\over R_5 R_6}\,,\spc i=1,2\,,
\ea
while those of the two D$(r;0,0,0)$-branes 
(cf. equations~(\ref{fracnormun}) and~(\ref{fracnormtw}))
\ba
{\cal N}^2_{f,U}&=&4\frac{V_{r+1}}{(2\pi)^{r+1}} {1\over 128}
{1\over\prod_{m=3}^8 R_m} \,,\\
{\cal N}^2_{f,Tg_1,i}&=&\frac{V_{r+1}}{(2\pi)^{r+1}} {2^4 \over 128}
{1\over R_3R_4}\,,\spc i=1,2\,,\\
{\cal N}^2_{f,Tg_3,i}&=&\frac{V_{r+1}}{(2\pi)^{r+1}} {2^4 \over 128}
{1\over R_5R_6}\,,\spc i=1,2\,.
\ea
The factor of 4 comes from the fact that there are two BPS branes.
These agree at the critical radius.

The second decay channel occurs when we decrease $R_7$ below
$1/\sqrt{2}$. The decay products are a pair of totally fractional
D$(r;0,0,2)$-branes. In the notation of section~\ref{eq43} one of the
D-branes has $\epsilon,\epsilon_1,\epsilon_3=+1$, and
$\theta_7,\theta_8=0$, while the other brane has
$\epsilon,\epsilon_1,\epsilon_3=-1$ and
$\theta_7,\theta_8=0$. Comparing the normalisation constants confirms
that charge and mass are conserved in the decay.

The third decay channel seems much more complicated. Naively one might
expect a decay into a brane-anti-brane combination of fractional
branes stretching along $x^8$ and one of
$x^3\,,x^4\,,x^5\,,x^6$. However, there are no such fractional
branes. If the dacay is to be into branes with two internal
directions, then by considering the Dirac quantisation condition these
will have to be fractional branes, which presumably will be somehow
bent away from the axes, and not have a boundary state description of
the type analysed in this paper.

It is also possible to analyse the stability of the
\dh$(r;0,0,1)$-brane charged only under,say, the $g_1$-twisted \RR
sector. Its stability region is
\be
R_5,R_6,R_7\ge\frac{1}{\sqrt{2}}\,,\spc\spc\spc\spc\;\; R_8\le\sqrt{2}\,,
\ee
and the $R_7$ and $R_8$ decay channels follow easily from the decay
channels of the \dh$(r;0,0,1)$-brane charged under the $g_1$ and
$g_2$-twisted \RR sectors discussed above.

\subsection{The \dh$(r;0,1,1)$-brane}

Finally, we discuss a \dh$(r;0,1,1)$-brane which we take to extend 
along $x^8$ and $x^6$, and be fixed
at $x^5=x^7=0$ and to consist of two identical copies at $(x^3,x^4)$
and $(-x^3,-x^4)$, as explained in appendix~\ref{appa}. The allowed boundary
states are of the form
\ba
\ket{{\hat D}(r;0,1,1),\varepsilon_i,\theta_j}&\!\!\!=&\!\!\!\!\!\!
\sum_{w_6,w_8} e^{i(\theta_6 w_6+\theta_8 w_8)}
\ket{{\hat D}(r;0,1,1),w_6,w_8}_{\mbox{\scriptsize \NSNS}}
\nonumber \\ & &\spc
\!\!\!\!\!\!\!\!\!+\,\varepsilon_1
\ket{{\hat D}(r;0,1,1)}_{\mbox{\scriptsize \RRT$g_{1,1}$}}
+\varepsilon_1e^{i\theta_6}
\ket{{\hat D}(r;0,1,1)}_{\mbox{\scriptsize \RRT$g_{1,2}$}}
\nonumber \\ & &\spc
\!\!\!\!\!\!\!\!\!+\,\varepsilon_1e^{i\theta_8}
\ket{{\hat D}(r;0,1,1)}_{\mbox{\scriptsize \RRT$g_{1,3}$}}
+\varepsilon_1e^{i(\theta_6+\theta_8)}
\ket{{\hat D}(r;0,1,1)}_{\mbox{\scriptsize \RRT$g_{1,4}$}}\,.
\ea
Hence, such truncated branes can only carry an even number of
negative twisted \RR charges. The stability of this brane
is very similar to those
encountered in $\Zop_2$ orbifolds, we simply state the results here.
It is stable for
\be
R_5,R_7\ge\frac{1}{\sqrt{2}}\,,\spc\spc\spc\spc\;\; R_6,R_8\le\sqrt{2}\,.
\ee
The \dh-brane can decay in two different ways: by increasing $R_6$  
or $R_8$ beyond
$\sqrt{2}$, or by decreasing $R_5$ or $R_7$ below $1/\sqrt{2}$.
In the first the decay is into a pair of singly charged
\dh$(r;0,0,1)$-branes, which carry the same charges as the
\dh$(r;0,1,1)$-brane. It is straightforward to check that at the
critical radius ($R_6=\sqrt{2}$) mass and charge are conserved.
The second decay channel corresponds to decreasing $R_5$ below
$1/\sqrt{2}$. The \dh-brane decays into a pair of singly charged
\dh$(r;0,2,1)$-branes, whose charges are the same at the four fixed
points on which the \dh$(r;0,1,1)$-brane ends and opposite at the
other four fixed points. Again both mass and charge are conserved at
the transition. The decay products of the \dh$(r;0,1,1)$-branes are
different from the decay products previously encountered. In
particular the brane does not decay into a brane-anti-brane pair of
separately BPS objects, rather the decay is into other non-BPS objects.

\section{T-duality}

In this section we discuss the T-dual of the $\Zop_2\times\Zop_2$
orbifold. We shall
T-dualise along the $x^5$ direction. The orbifold we consider
will then be a $\Zop_2\times\Zop_2$ orbifold where the generators are
$g_1=\I_{5678}(-1)^{F_l}$ and $g_2=\I_{3478}$, where $F_l$ is the left
moving space-time fermion number. One imposes GSO and
orbifold  invariance of
the boundary states in the different closed string sectors. As in the
$\I_4(-1)^{F_l}$ orbifolds there is an ambiguity regarding which
part of the twisted sector states to keep:  the $\I_4(-1)^{F_l}$ odd
states or the $\I_4(-1)^{F_l}$ even states? For the case of the
invariance under $g_i$ for the $g_i$-twisted sector we follow the
prescription of~\cite{S3}. In particular we require the $g_1$- and
$g_3$- twisted sectors to be odd under $g_1$ and $g_3$ respectively,
while the $g_2$-twisted sectors are to be even under $g_2$. We further
shall require the $g_1$-twisted sector to be $g_2$ even and $g_3$ odd,
the $g_2$-twisted sector to be $g_1$ and $g_3$ even and the
$g_3$-twisted sector to be $g_2$ even and $g_1$ odd. This ensures that
we can construct fractional branes. With these clarifications one
obtains table 1 for Type IIB

\begin{table}[hbt]
\begin{center}
\begin{tabular}{|c|c|c|c|c|} \hline
      & GSO             & $g_1$         & $g_2$          & $g_1g_2$   \\ \hline
\NSNS & --              & --            & --             & -- \\ 
\RR   & $r+s_1+s_2+s_3$ odd & $s_2+s_3$ odd & $s_1+s_3$ even & $s_1+s_2$ odd \\
\NSNST$g_1$&$s_2+s_3$ odd & $s_2+s_3$ odd & $s_3$ even  & $s_2$ odd \\
\RRT$g_1$&$r+s_1$ even & -- & $s_1$ even  & $s_1$ even \\
\NSNST$g_2$&$s_1+s_3$ even & $s_3$ even & $s_1+s_3$ even  & $s_1$ even \\
\RRT$g_2$&$r+s_2$ odd & $s_2$ odd & --  & $s_2$ odd \\
\NSNST$g_3$&$s_1+s_2$ odd & $s_2$ odd & $s_1$ even  & $s_1+s_2$ odd \\
\RRT$g_3$&$r+s_3$ even & $s_3$ even & $s_3$ even  & -- \\
\hline
\end{tabular}
\caption{Restrictions on $r$ and $s_i$ as a result of requiring GSO
and orbifold invariance for boundary states in the various closed
string sectors for Type IIB.}
\end{center}
\end{table}

With these restrictions it is not difficult to identify bulk,
fractional and both kinds of truncated branes. As before if two kinds
of branes can exist for a given $(r;\s)$it is the fractional
rather than truncated (or the truncated with the \NSNS twisted sector
rather than the other truncated brane) that will be fundamental and of
minimal charge. For Type IIB we then have bulk wrapped branes for
$r,s_1,s_3$ odd $s_2$ even (only charged under the untwisted \RR
sector), and bulk and fractional branes (latter charged under four
kinds of charges in the decompactified theory) for $r,s_1,s_3$ even
$s_2$ odd. Truncated branes charged under \RRT$g_1$ and
\RRT$g_2$ fields exist for $r,s_1$ even $s_2,s_3$ odd. Truncated
branes charged under \RRT$g_1$ and
\RRT$g_3$ fields exist for $r,s_1,s_2,s_3$ even. Truncated branes
charged under \RRT$g_2$ and
\RRT$g_3$ fields exist for $r,s_3$ even $s_1,s_2$ odd. Truncated
branes charged only under \RRT$g_1$ exist for $r,s_1,s_2$ even $s_3$
odd. Truncated
branes charged only under \RRT$g_2$ exist for $r$ even $s_1,s_2,s_3$
odd. Truncated
branes charged only under \RRT$g_3$ exist for $r,s_2,s_3$ even $s_1$
odd. For Type IIA in the above the parity of $r$ changes.

One can develop a K-theory understanding of this. We define a new
kind of K-theory which we call $K^*_{\Zop_2\times\Zop_2^\pm}$. Elements
of this are pairs of isomorphism classes of bundles $(E,F)$, which are
equivariant under the action of the first $\Zop_2$ and we are given an
isomorphism between $(E,F)$ and $(g^*(E),g^*(F))$ where in this case
$g=\I_{5678}$ and $g^*(E)$ is the pullback of $E$ by $g$. Further we
are given an isomorphism which maps $(E,F)$ to $(h^*(F),h^*(E))$,
where $h^*(E)$ is the pullback of $E$ by $h$, the generator of the
geometric part of the second $\Zop_2$ (in other words in this case
$h=\I_{3478}$). A version of Hopkins' formula then is
\be
K_{\Zop_2\times\Zop_2^{\pm}}^*(\Rop^{a;b,c,d})\cong
K_{\Zop_2\times\Zop_2}^*(\Rop^{a+1;b,c+1,d})\,,
\ee
where the second group is just the usual equivariant K-group. We have
computed the latter in section~\ref{sec3}, and using the above
relation exact agreement
between K-theory and the string results follows. The stability and
decay of the the truncated branes can easily be obtained from the
results of the T-dual scenario described in detail in the previous sections.

\appendix

\section{Construction and normalisation of boundary states}\label{appa}
\setcounter{equation}{0}

In this appendix we determine the normalisation constants of the
boundary states for the orbifold theories under consideration.

\subsection{The uncompactified case}\label{appa1}

In each (bosonic) sector of the theory we can construct the boundary
state 
\be\label{boundary1}
\ket{B(r;\s),k,\eta}=\exp\left(\sum_{l>0}^\infty 
\left[\frac{1}{l}\alpha_{-l}^\mu S_{\mu\nu}\tilde{\alpha}_{-l}^\nu\right]
+i\eta\sum_{m>0}^\infty \left[\psi_{-m}^\mu
S_{\mu\nu}\tilde{\psi}_{-m}^\nu\right]\right)
\ket{B(r;\s),k,\eta}^{(0)}\,,
\label{bdr}
\ee
where, depending on the sector, $l$ and $m$ are integer or
half-integer, and $k$ denotes the momentum of the ground state.  We
shall always work in light-cone gauge with light-cone directions $x^0$
and $x^9$; thus $\mu$ and $\nu$ take the values $1,\ldots, 8$. We
shall also drop the dependence on $\alpha^\prime$.

The parameter $\eta=\pm$ describes the two different spin structures 
\cite{PolCai,CLNY}, and the matrix $S$ encodes the boundary conditions
of the Dp-brane which we shall always take to be diagonal
\be
S=\mbox{diag}(-1,\dots,-1,1,\dots,1) \,,
\ee
where $p+1$ entries are equal to $-1$, $7-p$ entries are equal to
$+1$, and $p=r+s_1+s_2+s_3$. If there are fermionic zero modes, 
the ground state in (\ref{boundary1}) satisfies additional conditions
discussed in the following appendix. 

In order to obtain a localised D-brane, we have to take the Fourier
transform of the above boundary state, where we integrate over the
directions transverse to the brane,
\be\label{local}
\ket{B(r,\s),y,\eta}=
\int\left(\prod_{\mu \;\mbox{\scriptsize{transverse}}} 
dk^\mu e^{ik^\mu y_\mu}\right) dk^0e^{ik^0y_0} \;
dk^9 e^{ik^9 y_9} \; \ket{B(r;\s),k,\eta}\,,
\ee
$y$ denotes the location of the boundary state, and in the $g_i$-twisted
sector the momentum integral only involves transverse directions that
are not inverted by the action of $g_i$. In the following we shall
typically consider (without loss of generality) the case $y=0$ in which
case the boundary state is denoted by $\ket{B(r;\s),\eta}$.  

The invariance of the boundary state under the GSO-projection always
requires that the physical boundary state is a linear combination of
the two states corresponding to $\eta=\pm$. Using the conventions of
appendix~B in~\cite{S3}, these linear combinations are of the form
\ba
\left|B(r;\s)\right>_{\mbox{\scriptsize\NSNS}}&=&
\frac{1}{2}\Bigl(\left|B(r;\s),+\right>_{\mbox{\scriptsize\NSNS}}-
\left|B(r;\s),-\right>_{\mbox{\scriptsize\NSNS}}\Big)\,,\\
\left|B(r;\s)\right>_{\mbox{\scriptsize\RR}}&=&
2i\Bigl(\left|B(r;\s),+\right>_{\mbox{\scriptsize\RR}}+
\left|B(r;\s),-\right>_{\mbox{\scriptsize\RR}}\Bigr)\,,\\ 
\left|B(r;\s)\right>_{\mbox{\scriptsize\NSNST$g_i$}}&=&
\frac{1}{2}
\Bigl(\left|B(r;\s),+\right>_{\mbox{\scriptsize\NSNST$g_i$}}
+\left|B(r;\s),-\right>_{\mbox{\scriptsize\NSNST$g_i$}}\Bigr)\,,\\
\left|B(r;\s)\right>_{\mbox{\scriptsize\RRT$g_i$}}&=&
i\Bigl(\left|B(r;\s),+\right>_{\mbox{\scriptsize\RRT$g_i$}}
+\left|B(r;\s),-\right>_{\mbox{\scriptsize\RRT$g_i$}}\Bigr)
\,,\label{rrtw}
\ea
where, depending on the theory in question, these states are actually 
GSO-invariant provided that $r$ and $s_i$ satisfy suitable
conditions discussed in section~\ref{sec2} ($i=1,2,3$). The
normalisation constants have been 
introduced for later convenience.

In order to solve the open-closed consistency condition the actual
D-brane state is a linear combination of physical boundary states from 
different sectors. There are three elementary cases to consider, 
fully fractional, and the two truncated D-branes. (cf.
equations~\ref{fract})-(\ref{dbtr}) In 
the fully fractional case, the D-brane state can be written as 
\begin{eqnarray}
\ket{D(r;\s)} & = & {\cal N}_{f,U}
\left(\left|B(r;\s)\right>_{\mbox{\scriptsize\NSNS}} 
+ \epsilon \left|B(r;\s)\right>_{\mbox{\scriptsize\RR}} \right) 
\nonumber \\
& & \quad 
+ \sum_{i=1}^3 \epsilon_i \; {\cal N}_{f,Tg_i}
\left(\left|B(r;\s)\right>_{\mbox{\scriptsize\NSNST$g_i$}} 
+ \epsilon \left|B(r;\s)\right>_{\mbox{\scriptsize\RRT$g_i$}} \right)\,,
\end{eqnarray}
where $\epsilon=\pm$ determines the sign of the charge with respect
to the untwisted \RR sector charge, while $\epsilon_i=\pm$, $i=1,2,3$
determines the sign of the charge with respect to the \RRT$g_i$
charge. For consistency when $i\neq j\neq k\neq i$ we have
$\epsilon_i=\epsilon_j\epsilon_k$. The closed string cylinder diagram 
is then of the form
 \ba
\label{closedcal}
{\cal A} & = & \int dl \left<B(r;\s)\right|e^{-lH_c}
\left|B(r;\s)\right>
\nonumber \\ 
& = & \half {\cal N}^2_{f,U} \int dl \; l^{(p-9)/2}
\left(\frac{f_3^8(q)-f_4^8(q)-f_2^8(q)}{f_1^8(q)}\right) \nonumber \\
& & + \sum_{i=1}^3\half {\cal N}^2_{f,Tg_i} \int dl \; l^{(r+s_i-5)/2}
\left(\frac{f_3^{4}(q)f_2^4(q)-
f_2^{4}(q)f_3^4(q)}{f_1^{4}(q)f_4^4(q)}\right)\,,
\ea
where the functions $f_i$ are defined as in \cite{GabSen}, 
$q=e^{-2\pi l}$, and the closed string Hamiltonian is given by  
\be
H_c=\pi k^2+2\pi\sum_{\mu=1}^8\left[\sum_{l>0}^\infty
(\alpha^\mu_{-l}\alpha^\mu_l+\tilde{\alpha}^\mu_{-l}\tilde{\alpha}^\mu_l)+
\sum_{m>0}^\infty
m(\psi^\mu_{-m}\psi^\mu_m+\tilde{\psi}^\mu_{-m}\tilde{\psi}^\mu_m)\right]
+2\pi C_c\,.
\ee
Here the constant $C_c$ is $-1$ in the \NSNS sector, and zero in all
other sectors. The corresponding open string amplitude is obtained by the 
modular transformation $t=1/2l$, ${\tilde q}=e^{-\pi t}$,
\ba
{\cal A} & = & 
2^{(7-p)/2} {\cal N}^2_{f,U} \int {dt \over 2t}
t^{-(p+1)/2} 
\left(\frac{f_3^8(\tilde{q})-f_2^8(\tilde{q})
-f_4^8(\tilde{q})}{f_1^8(\tilde{q})}\right) \nonumber \\
& & + \sum_{i=1}^32^{(3-r-s_i)/2} {\cal N}^2_{f,Tg_i} \int {dt \over 2t}
t^{-(r+s_i+1)/2} \left(\frac{f_3^{4}({\tilde q})f_4^4({\tilde q})-
f_4^{4}({\tilde q})f_3^4({\tilde q})}
{f_1^{4}({\tilde q}) f_2^4({\tilde q})}\right)\,.
\label{theclosedone}
\ea
This is to be compared with the open string one-loop diagram,
\ba
\int\frac{dt}{2t}&
&\!\!\!\!\!\!\!\!\!\!\!\!\!\!\!
\Tr_{NS-R}\left(\frac{1+(-1)^F}{2}\frac{1+g_1+g_2+g_3}{4}
e^{-2tH_o}\right)\nonumber \\
&=&
\frac{V_{p+1}}{(2\pi)^{p+1}}2^{-(p+7)/2}\int
\frac{dt}{2t}t^{-(p+1)/2}
\left(\frac{f_3^8({\tilde q})-f_4^8({\tilde q})-
f_2^8({\tilde q})}{f_1^8({\tilde q})}\right)
\nonumber \\
& &+\sum_{i=1}^3\frac{V_{r+s_i+1}}{(2\pi)^{r+s_i+1}}2^{-(r+s_i+3)/2}\int
\frac{dt}{2t}t^{-(r+s_i+1)/2}
\left(\frac{f_3^{4}({\tilde q})f_4^4({\tilde q})
-f_4^{4}({\tilde q})f_3^4({\tilde q})}
{f_1^{4}({\tilde q})f_2^4({\tilde q})}\right)\,,\nonumber\\
\label{theopenone}
\ea
where $V_{p+1}$ is the
(infinite) $p+1$ dimensional volume of the brane, whilst
$V_{r+s_i+1}$ is the volume of the projection onto the directions
unaffected by $g_i$. The open string Hamiltonian is given by
\be
H_o=\pi p^2+\pi\sum_{\mu=1}^8\left[\sum_{l>0}^\infty \alpha_{-l}^\mu 
\alpha_l^\mu+\sum_{m>0}^\infty m\psi^\mu_{-m}\psi^\mu_m\right] +\pi
C_o\,, 
\ee
where, in the R sector, $l$ and $m$ run over the positive integers for
NN and DD directions, and over positive half integers for ND
directions. In the NS sector, the moding of the fermions (and
therefore the values for $m$) are opposite to those in the R sector.
$C_o$ is zero in the R sector and is $\frac{4-t}{8}$ in the NS sector,
where $t$ is the number of ND directions. Comparison of
equations~(\ref{theopenone}) and~(\ref{theclosedone}) then gives
\ba
{\cal N}^2_{f,U}&=&\frac{V_{p+1}}{(2\pi)^{p+1}}
{1\over 128}\,, \\
{\cal N}^2_{f,Tg_i}&=&\frac{V_{r+s_i+1}}{(2\pi)^{r+s_i+1}}
{1 \over 8}\,.\label{con1}
\ea

Consider next the singly fractional D-branes charged under the
\RRT$g_i$ field. These can be thought of
as superpositions of two totally fractional D-branes with opposite
twisted charges in the other two twisted sectors. These pairs can move
off the fixed points of $g_j$ for $j\neq i$, such that they lie at
positions $x^\mu$ and $-x^\mu$ for the directions transverse to the
brane fixed by $g_j$ and not by $g_i$. In particular their
boundary states will now look like 
\begin{eqnarray}
\ket{D(r;\s)} &\!\!\! = &\!\!\! {\cal N}_{f,U}
\left(\left|B(r;\s),x^\mu\right>_{\mbox{\scriptsize\NSNS}}
\!+\left|B(r;\s),-x^\mu\right>_{\mbox{\scriptsize\NSNS}}\right.
\nonumber \\
& & \left.\quad
\!+ \epsilon( \left|B(r;\s),x^\mu\right>_{\mbox{\scriptsize\RR}}
\!+ \left|B(r;\s),-x^\mu\right>_{\mbox{\scriptsize\RR}}) 
\right) 
\nonumber \\
& & \quad 
+\epsilon_i \; {\cal N}_{f,Tg_i}
\left(\left|B(r;\s),x^\mu\right>_{\mbox{\scriptsize\NSNST$g_i$}} 
+\left|B(r;\s),-x^\mu\right>_{\mbox{\scriptsize\NSNST$g_i$}}\right.
\nonumber \\
& &\quad 
\left.
+ \epsilon \left|B(r;\s),x^\mu\right>_{\mbox{\scriptsize\RRT$g_i$}}
+ \epsilon
\left|B(r;\s),-x^\mu\right>_{\mbox{\scriptsize\RRT$g_i$}} 
\right)\,.
\end{eqnarray}
The above normalisation is consistent with the fact that the cylinder 
diagram for such D-branes factorises on 
\be
2\int\frac{dt}{2t}
\Tr_{NS-R}\left(\frac{1+(-1)^F}{2}\frac{1+g_i}{2}e^{-2tH_o}\right)\,.
\ee
Such singly fractional D-branes correspond to, in the covering space,
two Type II D-branes stuck at the fixed point of $g_i$. There are four
kinds of
open strings stretching between two such D-branes so one would expect
a factor of four in front of the trace above. However, the orbifold
identifies certain strings leaving only two independent ones (namely
the string that has endpoint on the same brane and the string that has
ends on opposite branes). Similarly a bulk D-brane is a configuration
of four totally fractional D-branes whose twisted charges cancel. It's
boundary state is a sum over four \NSNS and four \RR 
boundary states at positions mapped to one another under the orbifold
action. The cylinder diagram for such a D-brane factorises on 
\be
4\int\frac{dt}{2t}
\Tr_{NS-R}\left(\frac{1+(-1)^F}{2}e^{-2tH_o}\right)\,.
\ee

The analysis for the case of the truncated \dh-branes is similar. 
The truncated D-brane of the type given by
equation~(\ref{sgtr}) charged only under the \RRT$g_1$ sector, say, has a 
boundary state 
\ba\label{singtrunc}
\ket{\hat{D}(r;\s)} &=& {\cal N}_{ts,U}\left(
\left|B(r;\s),x^\mu\right>_{\mbox{\scriptsize\NSNS}} 
+\left|B(r;\s),-x^\mu\right>_{\mbox{\scriptsize\NSNS}}\right)\nonumber
\\& &\quad
+ \epsilon {\cal N}_{ts,Tg_1}\left(
\left|B(r;\s),x^\mu\right>_{\mbox{\scriptsize\RRT$g_1$}}
+\left|B(r;\s),-x^\mu\right>_{\mbox{\scriptsize\RRT$g_1$}}\right)\,,
\ea
where $\epsilon=\pm$ determines the sign of the \RRT$g_1$ sector
charge and $x^\mu$ denotes some of the directions $x^3,x^4$
which are transverse to the
\dh-brane. The closed string tree diagram now only produces some of the
terms of (\ref{closedcal}), and the corresponding open string
amplitude is 
\ba
\label{singtruncann}
4\int\frac{dt}{2t}&
&\!\!\!\!\!\!\!\!\!\!\!\!\!\!\!
\Tr_{NS-R}\left(\frac{1+ g_1(-1)^F}{2}
e^{-2tH_o}\right)\nonumber \\
&=&
4\frac{V_{p+1}}{(2\pi)^{p+1}}2^{-(p+3)/2}\int
\frac{dt}{2t}t^{-(p+1)/2}
\left(\frac{f_3^8({\tilde q})-f_2^8({\tilde q})}
{f_1^8({\tilde q})}\right) 
\nonumber \\
& & - 4\frac{V_{r+s_1+1}}{(2\pi)^{r+s_1+1}}2^{(1-s_1-r)/2}\int
\frac{dt}{2t}t^{-(r+s_1+1)/2}
\left(\frac{f_4^{4}({\tilde q})f_3^4({\tilde q})}
{f_1^{4}({\tilde q})f_2^4({\tilde q})}\right)\,.
\ea
Comparison with the corresponding closed string calculation then gives
\ba
{\cal N}^2_{t1,U}&=&\frac{V_{p+1}}{(2\pi)^{p+1}}{1 \over 8}\,, \\
{\cal N}^2_{t1,Tg_1}&=&\frac{V_{r+s_1+1}}{(2\pi)^{r+s_1+1}}2\,. 
\label{con2} 
\ea

The truncated D-brane of the type given in equation~(\ref{dbtr}) is
charged under the \RRT$g_i$ and \RRT$g_j$
sectors $(i\neq j)$ has a boundary state given by
\ba
\ket{\hat{D}(r;\s)} &=& {\cal N}_{td,U}
\left|B(r;\s)\right>_{\mbox{\scriptsize\NSNS}} 
+ \epsilon_i {\cal N}_{td,Tg_i}
\left|B(r;\s)\right>_{\mbox{\scriptsize\RRT$g_i$}}\nonumber \\
& &\quad +\; \epsilon_j {\cal N}_{td,Tg_j}
\left|B(r;\s)\right>_{\mbox{\scriptsize\RRT$g_j$}}
+ \epsilon_i\epsilon_j {\cal N}_{td,Tg_k}
\left|B(r;\s)\right>_{\mbox{\scriptsize\NSNST$g_k$}}\, 
\ea
where $\epsilon_i=\pm$ determines the sign of the \RRT$g_i$ sector
charge and $j\neq k\neq i$ with $k=1,2,3$. 
The closed string tree diagram produces only some of the
terms of (\ref{closedcal}), and the corresponding open string
amplitude is 
\ba
\int\frac{dt}{2t}&
&\!\!\!\!\!\!\!\!\!\!\!\!\!\!\!
\Tr_{NS-R}\left(\frac{1+g_i(-1)^F}{2}\frac{1+g_j(-1)^F}{2}
e^{-2tH_o}\right)\nonumber \\
&=&
\frac{V_{p+1}}{(2\pi)^{p+1}}2^{-(p+5)/2}\int
\frac{dt}{2t}t^{-(p+1)/2}
\left(\frac{f_3^8({\tilde q})-f_2^8({\tilde q})}
{f_1^8({\tilde q})}\right) 
\nonumber \\
& & -\sum_{\alpha\in\{i,j,k\}}\rho(\alpha)
\frac{V_{r+s_\alpha+1}}{(2\pi)^{r+s_\alpha+1}}
2^{-(r+s_\alpha+1)/2}\int
\frac{dt}{2t}t^{-(r+s_\alpha+1)/2}
\left(\frac{f_4^{4}({\tilde q})f_3^4({\tilde q})}
{f_1^{4}({\tilde q})f_2^4({\tilde q})}\right)\,,
\ea
where $\rho(i)=\rho(j)=-\rho(k)=1$. Comparison with the corresponding 
closed string calculation then gives
\ba
{\cal N}^2_{t2,U}&=&\frac{V_{p+1}}{(2\pi)^{p+1}}{1 \over 64}\,, \\
{\cal N}^2_{t2,Tg_\alpha}&=&
\frac{V_{r+s_\alpha+1}}{(2\pi)^{r+s_\alpha+1}}{1\over 4}\,. 
\label{con3} 
\ea

\subsection{The compactified case}

The construction in the compactified case is essentially the same as
in the above uncompactified case; however there are the following
differences. 

\noindent 1. In the localised boundary state (\ref{local}) the
integral over compact transverse directions is replaced by a sum
\be
\int dk^\nu e^{ik^\nu y_\nu} \longrightarrow
\sum_{m^\nu\in\Zop} e^{i m^\nu y_\nu / R_\nu} \,,
\ee
where $R_\nu$ is the radius of the compact $x^\nu$ direction.

\noindent 2. In the two untwisted sectors, the ground state is
in addition characterised by a winding number $w_\nu$ for each compact 
direction that is tangential to the world-volume of the brane. In the
$g_i$ twisted sectors, the ground states will also be characterised by
winding numbers in each of the $s_i$ directions tangential to the
world-volume of the brane. The
localised bound state (\ref{local}) then also contains a sum over these
winding states 
\be
\sum_{w_\mu} e^{i\theta^\mu w_\mu} \,,
\ee
where $\theta^\mu$ is a Wilson line; as required by orbifold
invariance, $\theta^\mu \in\{ 0,\pi\}$. 

\noindent 3. For general $s_i$, the contribution in the twisted
sectors consists of a sum of terms that are associated to $2^{s_i}$ of the 
$16$ different twisted sectors that define the endpoints of the
world-volume of the brane in the internal space. For convenience we
may assume that one of the $2^{s_i}$ fixed points is always the origin.   

\noindent 4. The open and closed string Hamiltonians, $H_o$ and $H_c$,
each acquire an extra term $1/4\pi(\sum_\mu w_\mu^2)$.

Let us now construct in more detail the boundary state for a fully 
fractional D$(r;\s)$ brane. This is of the form
\begin{eqnarray}\label{fractboun}
\ket{D(r;\s)} &\!\!\!\! = &\!\!\!\! {\cal N}_{f,U}
\left(\left|B(r;\s)\right>_{\mbox{\scriptsize\NSNS}} 
+ \varepsilon \left|B(r;\s)\right>_{\mbox{\scriptsize\RR}} \right) 
\nonumber \\
& & +\sum_{i=1}^3
 \varepsilon_i \; {\cal N}_{f,Tg_i} \sum_{\alpha_i=1}^{2^{s_i}}
\!e^{i\theta_{\alpha_i}}\!
\left(\left|B(r;\s)\right>_{\mbox{\scriptsize\NSNST$g_{\alpha_i}$}} 
\!+ \!\epsilon 
\left|B(r;\s)\right>_{\mbox{\scriptsize\RRT$g_{\alpha_i}$}}
\right)\,,
\end{eqnarray}
where $\alpha_i$ labels the different fixed points between which the
brane stretches (where we choose the convention that T$g_{i_1}$ is the
twisted sector at the origin), and $\theta_{\alpha_i}$ is the Wilson
line that is associated to the difference of the fixed point $\alpha_i$
and the origin. The closed string tree diagram is now
\ba
{\cal A}_c &=&  \int dl \left<B(r;\s)\right|e^{-lH_c}
\left|B(r;\s)\right>
\nonumber \\ 
& = & \half{\cal N}^2_{f,U} \int dl \;
l^{(r-3)/2}
\left(\frac{f_3^8(q)-f_2^8(q)-f_4^8(q)}{f_1^8(q)}\right)
\nonumber \\
& & \qquad \qquad\times\;\;\
\prod_{m=1}^{s_1+s_2+s_3}\sum_{w_{j_m}\in\Zop} e^{-l\pi R_{j_m}^2 w_{j_m}^2}
\prod_{m=1}^{6-s_1-s_2-s_3}\sum_{n_{k_m}\in\Zop} e^{-l\pi (n_{k_m}/R_{k_m})^2}
\nonumber \\
& & \quad +\sum_{i=1}^3 {2^{s_1+s_2+s_3-s_i} \over 2} {\cal N}^2_{f,Tg_i}
\int dl \;
l^{(r-3)/2}
\left(\frac{f_3^{4}(q)f_2^4(q)-
f_2^{4}(q)f_3^4(q)}{f_1^{4}(q)f_4^4(q)}\right)
\nonumber \\
& & \qquad \qquad\times\;\;\
\prod_{m=1}^{s_i}\sum_{w_{j_m}\in\Zop} e^{-l\pi R_{j_m}^2 w_{j_m}^2}
\prod_{m=1}^{2-s_i}\sum_{n_{k_m}\in\Zop} e^{-l\pi (n_{k_m}/R_{k_m})^2} 
\,,
\ea
where $R_{j_m}$, $m=1,\ldots, s_1+s_2+s_3$ 
are the radii of the circles that are
tangential to the world-volume of the brane, and $R_{k_m}$, 
$i=1,\ldots, 6-s_1-s_2-s_3$ are the radii of the directions transverse to the
brane. Upon the substitution $t=1/2l$, using the Poisson resummation
formula (see for example~\cite{sen2,GabSen}), this amplitude becomes
\ba
{\cal A}_c &=& {\cal N}^2_{f,U} 
{\prod_{m=1}^{6-s_1-s_2-s_3} R_{k_m} \over \prod_{m=1}^{s_1+s_2+s_3}  R_{j_m}} 
2^{(7-r)/2} \int {dt \over 2t} t^{-(r+1)/2} 
\left(\frac{f_3^8(\tilde{q})-f_2^8(\tilde{q})
-f_4^8(\tilde{q})}{f_1^8(\tilde{q})}\right) \nonumber \\
& & \qquad \qquad\times\;\;\ 
\prod_{m=1}^{s_1+s_2+s_3}\sum_{n_{j_m}\in\Zop} e^{-2t\pi n_{j_m}^2 / R_{j_m}^2}
\prod_{m=1}^{6-s_1-s_2-s_3}\sum_{w_{k_m}\in\Zop} e^{-2t\pi w_{k_m}^2 R_{k_m}^2}
\nonumber \\
& & \quad + \sum_{i=1}^3 
{\prod_{m=1}^{2-s_i} R_{k_m} \over \prod_{m=1}^{s_i}  R_{j_m}}
2^{(1-r)/2} 2^{s_1+s_2+s_3-s_i} {\cal N}^2_{f,Tg_i} 
\int {dt \over 2t}
t^{-(r+1)/2} \left(\frac{f_3^{4}({\tilde q})f_4^4({\tilde q})-
f_4^{4}({\tilde q})f_3^4({\tilde q})}
{f_1^{4}({\tilde q}) f_2^4({\tilde q})}\right) \nonumber \\
& & \qquad \qquad\times\;\;\ 
\prod_{m=1}^{s_i}\sum_{n_{j_m}\in\Zop} e^{-2t\pi n_{j_m}^2 / R_{j_m}^2}
\prod_{m=1}^{2-s_i}\sum_{w_{k_m}\in\Zop} e^{-2t\pi w_{k_m}^2 R_{k_m}^2}\,.
\nonumber 
\ea
This is to be compared with the open string amplitude
\ba
\int\frac{dt}{2t}
& &\Tr_{NS-R}\left(\frac{1+(-1)^F}{2}\frac{1+g_1+g_2+g_3}{4}
e^{-2tH_o}\right) \nonumber \\
& = & \frac{V_{r+1}}{8 (2\pi)^{r+1}} 2^{-(r+1)/2}
\int\frac{dt}{2t}\! t^{-(r+1)/2}
\left(\frac{f_3^8({\tilde{q}})-f_4^8({\tilde q})-
f_2^8({\tilde q})}{f_1^8({\tilde q})}\right)
\nonumber \\
& &\qquad\times\;
\prod_{m=1}^{s_1+s_2+s_3}\sum_{n_{j_m}\in\Zop} e^{-2t\pi n_{j_m}^2 / R_{j_m}^2}
\prod_{m=1}^{6-s_1-s_2-s_3}\sum_{w_{k_m}\in\Zop} e^{-2t\pi w_{k_m}^2 R_{k_m}^2}
\nonumber \\
& &  \sum_{i=1}^3 \frac{V_{r+1}}{2(2\pi)^{r+1}}2^{-(r+1)/2}\int
\frac{dt}{2t}t^{-(r+1)/2} 
\frac{f_3^{4}({\tilde q})f_4^4({\tilde q})
-f_4^{4}({\tilde q})f_3^4({\tilde q})}
{f_1^4({\tilde q})f_2^4({\tilde q})}
\nonumber \\
& &\qquad\times\;
\prod_{m=1}^{s_i}\sum_{n_{j_m}\in\Zop} e^{-2t\pi n_{j_m}^2 / R_{j_m}^2}
\prod_{m=1}^{2-s_i}\sum_{w_{k_m}\in\Zop} e^{-2t\pi w_{k_m}^2 R_{k_m}^2}
\,.
\ea
By comparison this then fixes the normalisation constants as
\ba
{\cal N}^2_{f,U}&=&\frac{V_{r+1}}{(2\pi)^{r+1}} {1\over 128}
{\prod_{m=1}^{s_1+s_2+s_3} R_{j_m}\over\prod_{m=1}^{6-s_1-s_2-s_3} R_{k_m}} \,,
\label{fracnormun}\\
{\cal N}^2_{f,Tg_i}&=&\frac{V_{r+1}}{(2\pi)^{r+1}} {2^{4-(s_1+s_2+s_3-s_i)} \over 128}
{\prod_{m=1}^{s_i} R_{j_m}\over\prod_{m=1}^{2-s_i} R_{k_m}}\,.
\label{fracnormtw}
\ea
The extension of this to the singly fractional and bulk D-branes is
obvious.

The analysis for the truncated D-branes is almost identical. The
boundary state of a truncated D-brane of the type given in
equation~(\ref{sgtr}) is the truncation of
(\ref{fractboun}) to the 
untwisted \NSNS and the $g_i$-twisted \RR sectors with the addition of
boundary states at mirror positions as in equation~(\ref{singtrunc}). 
The open string amplitude contains also only 
the corresponding terms. Furthermore, since the projection operator is 
now\footnote{cf equation~(\ref{singtruncann})}
$4\times\half(1+g_i(-1)^F)$ each of the terms that 
appears is sixteen times as large as in the fractional case above. 
This implies that the relevant normalisation constants are given as 
\ba
{\cal N}^2_{t1,U}&=&\frac{V_{r+1}}{(2\pi)^{r+1}} {1\over 8}
{\prod_{m=1}^{s_1+s_2+s_3} R_{j_m}\over\prod_{m=1}^{6-s_1-s_2-s_3} R_{k_m}} \,,
\label{singtruncnormun}\\
{\cal N}^2_{t1,Tg_i}&=&\frac{V_{r+1}}{(2\pi)^{r+1}} 
{2^{4-(s_1+s_2+s_3-s_i)} \over 8}
{\prod_{m=1}^{s_i} R_{j_m}\over\prod_{m=1}^{2-s_i} R_{k_m}}\,.
\label{singtruncnormtw}
\ea
The second type of truncated \dh-brane (equation~(\ref{dbtr})) contains
the untwisted \NSNS, twisted
\RRT$g_i$, \RRT$g_j$ and \NSNST$g_k$ boundary states. The
projection operator is ${1\over 4}(1+g_i(-1)^F)(1+g_j(-1)^F)$ thus
fixing the normalisations to be
\ba
{\cal N}^2_{t2,U}&=&\frac{V_{r+1}}{(2\pi)^{r+1}} {1\over 64}
{\prod_{m=1}^{s_1+s_2+s_3} R_{j_m}\over\prod_{m=1}^{6-s_1-s_2-s_3} R_{k_m}} \,,
\label{dbtruncnormun}\\
{\cal N}^2_{t2,Tg_\alpha}&=&\frac{V_{r+1}}{(2\pi)^{r+1}} {2^{4-(s_1+s_2+s_3-s_\alpha)} \over 64}
{\prod_{m=1}^{s_\alpha} R_{j_m}\over\prod_{m=1}^{2-s_\alpha} R_{k_m}}\,,
\label{dbtruncnormtw}
\ea
where $\alpha\in\{i,j,k\}$.

\section{Consistency conditions of boundary states}\label{appb}
\setcounter{equation}{0}

In this appendix we discuss the invariance of the boundary states under
the GSO and orbifold actions.
Since the action of the orbifold group is purely geometrical the 
GSO projection is the same in twisted and untwisted sectors namely we have
\ba
\mbox{\NSNS}\spc\frac{1}{4}(1+(-1)^F)(1+(-1)^{{\tilde F}}) \,,\\
\mbox{\RR}\spc\frac{1}{4}(1+(-1)^F)(1\mp(-1)^{{\tilde F}}) \,,
\ea
for Type IIA and IIB, respectively. 

The GSO invariance of each of the sectors' boundary states was 
computed in detail in~\cite{S3}. Here we simply restate those results in
terms of $r$ and $s_i$. We shall work in the light-cone gauge with the
directions $x^0,x^9$ always Dirichlet~\cite{GrGut}. 
In the untwisted \NSNS sector it is easy to see that
\be
\ket{B(r;\s)}_{\mbox{\scriptsize\NSNS}}=\frac{{\cal N}}{2}
(\ket{B(r;\s),+}_{\mbox{\scriptsize\NSNS}}-
\ket{B(r;\s),-}_{\mbox{\scriptsize\NSNS}})
\ee
is GSO invariant for all $r,s_i$. The exact values of the
normalisation constants of the boundary states will depend on the kind
of D-brane the boundary state is a part of, and are computed in the
Appendix. In the untwisted \RR sector 
\be
\ket{B(r;\s)}_{\mbox{\scriptsize\RR}}=\frac{4i{\cal N}}{2}
(\ket{B(r;\s),+}_{\mbox{\scriptsize\RR}}
+\ket{B(r;\s),-}_{\mbox{\scriptsize\RR}})\,
\ee
is a GSO invariant boundary state if $r+s_1+s_2+s_3$ is even/odd for Type
IIA/B, respectively. For the \NSNST$g_1$ sector we find that
\be
\ket{B(r;\s)}_{\mbox{\scriptsize\NSNST$g_1$}}=
{\cal N}_T(\ket{B(r;\s),+}_{\mbox{\scriptsize\NSNST$g_1$}}
+\ket{B(r;\s),-}_{\mbox{\scriptsize\NSNST$g_1$}})\,
\ee
is a GSO invariant boundary state provided $s_2+s_3$ is even, while
\be
\ket{B(r;\s)}_{\mbox{\scriptsize\RRT$g_1$}}=
i{\cal N}_T(\ket{B(r;\s),+}_{\mbox{\scriptsize\RRT$g_1$}}
+\ket{B(r;\s),-}_{\mbox{\scriptsize\RRT$g_1$}})\,
\ee
is GSO invariant for $r+s_1$ even/odd for Type IIA/IIB. Similarly
in the \NSNST$g_2$ sector $s_1+s_3$ is to be even and for the \RRT$g_2$
sector 
$r+s_2$ has to be even/odd for Type IIA/IIB. Finally for the \NSNST$g_3$ 
$s_1+s_2$ is to be even and for the \RRT$g_2$ sector $r+s_3$ has to be
even/odd for Type IIA/IIB.

Next one requires the boundary states to be invariant under $g_i$. As
usual~\cite{S3} this places no restrictions on the untwisted \NSNS sector, and 
in the untwisted \RR sector it requires for $s_1+s_2$, $s_2+s_3$ and
$s_1+s_3$ to be all even. The \RRT$g_i$ 
boundary state has no restrictions placed on it by $g_i$ since it has
no zero modes in the directions which $g_i$ inverts, while the 
\NSNST$g_i$ boundary state's invariance under $g_i$ is equivalent to 
the GSO condition on that boundary state~\cite{S3}.

New non-trivial restrictions arise by requiring the $g_i$-twisted
sector's boundary state be invariant under $g_j$, $j\neq i$. We consider
the invariance of \NSNST$g_1$ and \RRT$g_1$ under $g_2$ in some detail.
The other conditions will follow in a similar way. Since the \NSNST$g_1$
sector has zero modes in the directions $x^7$ and $x^8$ (as well as $x^5$
and $x^6$) $g_2$ will have a non-trivial representation on these
zero-modes, given by\footnote{We use the same convention here for $g_2$ as
was used for ${\cal I}_2$ in~\cite{S3}. Since this is a supersymmetric
theory this can be viewed as a condition for supersymmetry
preservation by the orbifold action.} 
\be
g_2=\prod_{\mu=7,8}(\sqrt{2}\psi^\mu_0)
\prod_{\mu=7,8}(\sqrt{2}{\tilde\psi}^\mu_0)\,. 
\ee 
\ni This operator squares to one. It is not difficult to see that 
\be
g_2\ket{B(r;\s),\eta}^0_{\mbox{\scriptsize \NSNST$g_1$}}=
(-1)^{s_3}\ket{B(r;\s),\eta}^0_{\mbox{\scriptsize \NSNST$g_1$}} 
\ee 
\ni and hence that $s_3$ has to be even.\footnote{Note that in the notation 
of equations (B.4) and (B.5) Appendix B in~\cite{S3} for the
\NSNST$g_1$ state $a=b=1$.}
Similarly since the \RRT$g_2$ sector has zero modes in directions $x^3$
and $x^4$, $g_2$ has a non-trivial representation on this sector as 
\be
g_2=\prod_{\mu=3,4}(\sqrt{2}\psi^\mu_0)
\prod_{\mu=3,4}(\sqrt{2}{\tilde\psi}^\mu_0)\,, 
\ee 
and so 
\be
g_2\ket{B(r;\s),\eta}^0_{\mbox{\scriptsize \RRT$g_1$}}=
(-1)^{s_1}
\ket{B(r;\s),\eta}^0_{\mbox{\scriptsize \RRT$g_1$}}\,. 
\ee 
Thus $s_1$ has to be even.\footnote{Following equations (B.12) and (B.13)
of~\cite{S3} $\hat{a}=\hat{b}=1$ for the \RRT$g_1$ ground state.}
Performing a similar analysis for the other twisted sectors one
finds that in Type IIB the following boundary states are GSO and
orbifold invariant
\ba 
\ket{B(r;\s)}_{\mbox{\scriptsize\NSNS}}& &\mbox{for all
$r$ and $s_i$,} \nonumber \\ 
\ket{B(r;\s)}_{\mbox{\scriptsize\RR}}& &\mbox{for either $r$ odd and $s_i$
all even or $r$ even and $s_i$ all odd,} \nonumber \\ 
\ket{B(r;\s)}_{\mbox{\scriptsize\NSNST$g_i$}}& &\mbox{
for $s_j,s_k$ even, $j,k\neq i$,}\nonumber \\ 
\ket{B(r;\s)}_{\mbox{\scriptsize\RRT$g_i$}}& &\mbox{for $r$ odd and $s_i$
even.} \nonumber
\ea 
For Type IIA the conditions are the same for $s_i$ but $r$ has to be even.

\section*{Acknowledgments} 

I am grateful to D.-E. Diaconescu, M.B. Green, F.Quevedo, A. Sen,
B. Totaro and in
particular to M.R. Gaberdiel and G. Segal for many helpful discussions
and insights. B.S. is supported by the Cambridge Commonwealth Trust.

\ed